\documentclass[letter]{jpsj2} 
\title{Weak Magnetic Order in the Bilayered-hydrate Na$_{x}$CoO$_{2}\cdot y$H$_{2}$O Structure Probed by Co Nuclear Quadrupole Resonance\\
--- Proposed Phase Diagram in Superconducting  Na$_x$CoO$_{2} \cdot$ $y$H$_2$O ---  }
\author{
Y. \textsc{Ihara,}$^{1}$\thanks{E-mail address: ihara@scphys.kyoto-u.ac.jp}
K. \textsc{Ishida,}$^{1}$\thanks{E-mail address: kishida@scphys.kyoto-u.ac.jp}
C. \textsc{Michioka,}$^{2}$
M. \textsc{Kato,}$^{2}$
K. \textsc{Yoshimura,}$^{2}$
K. \textsc{Takada,}$^{3}$
T. \textsc{Sasaki,}$^{3}$
H. \textsc{Sakurai,}$^{4}$
and
E. \textsc{Takayama-Muromachi}$^{4}$}

\inst{$^{1}$Department of Physics, Graduate School of Science, Kyoto University, Kyoto 606-8502, Japan. \\
$^{2}$Department of Chemistry, Graduate School of Science, Kyoto University, Kyoto 606-8502, Japan. \\
$^{3}$Advanced Materials Laboratory, National Institute for Materials Science, 1-1 Namiki, Tsukuba, Ibaraki, 305-0044, Japan.\\
$^{4}$Superconducting Materials Center, National Institute for Materials Science, 1-1 Namiki, Tsukuba, Ibaraki 305-0044, Japan.}

\abst{
A weak magnetic order was found in a non-superconducting bilayered-hydrate Na$_{x}$CoO$_{2}\cdot y$H$_{2}$O sample by a Co Nuclear Quadrupole Resonance (NQR) measurement.
The nuclear spin-lattice relaxation rate divided by temperature $1/T_1T$ shows a prominent peak at 5.5 K, below which a Co-NQR peak splits due to an internal field at the Co site.  
From analyses of the Co NQR spectrum at 1.5 K, the internal field is evaluated to be $\sim$ 300 Oe and is in the $ab$-plane.
The magnitude of the internal field suggests that the ordered moment is as small as $\sim 0.015$ $\mu_B$ using the hyperfine coupling constant reported previously.
It is shown that the NQR frequency $\nu_Q$ correlates with magnetic fluctuations from measurements of NQR spectra and $1/T_1T$ in various samples. The higher-$\nu_Q$ sample has the stronger magnetic fluctuations.
A possible phase diagram in  Na$_{x}$CoO$_{2}\cdot y$H$_{2}$O is depicted using $T_c$ and $\nu_Q$, in which the crystal distortion along the $c$-axis of the tilted CoO$_2$ octahedron is considered to be a physical parameter.
Superconductivity with the highest $T_c$ is seemingly observed in the vicinity of the magnetic phase, suggesting strongly that the magnetic fluctuations play an important role for the occurrence of the superconductivity.

}

\kword{superconductivity, hydrate sodium cobalt oxide, NQR, spin fluctuations}

\begin{document}
\maketitle

Since the discovery of superconductivity in the bilayered-hydrate (BLH) Na$_{x}$CoO$_{2}\cdot y$H$_{2}$O (NCO) by Takada {\it et al.} \cite{takada}, a lot of works have been performed to investigate physical properties of this superconductivity.
Different from the CuO$_{2}$ square lattice in cuprate superconductors, the CoO$_{2}$ plane, in which superconductivity in BLH NCO occurs, is formed by a triangular lattice.
The geometrical frustrations inherent to the triangular lattice have been invoked to play a role for the occurrence of the superconductivity.

The crystal structure of BLH NCO consists of double hydrate layers sandwiching a Na layer and two-dimensional CoO$_{2}$ block layers, in which CoO$_6$ octahedra are tilted and connect each other by edge sharing.
In the presence of the distortion along the $c$-axis in the tilted CoO$_{6}$ octahedron, three degenerated $t_{2g}$ orbitals are lifted and split into doublet $e'_{g}$ orbitals and a singlet $a_{1g}$ orbital.
From a theoretical point of view, it is suggested that $e'_{g}$ character which forms six hole-like Fermi surface near $K$ point plays an essential role for the occurrence of the unconventional superconductivity.\cite{Ikeda,Yanase,Mochizuki}
However, these hole pockets are not observed by angle-resolved photo-emission spectroscopy (ARPES) studies in Na$_{x}$CoO$_{2}$ without water.\cite{ARPES1,ARPES2}
It is a crucial issue to experimentally determine which orbitals, the $e'_{g}$ orbitals or $a_{1g}$ orbital, are active for the superconductivity in BLH NCO.

Quite recently, a non-superconducting (SC) sample with BLH structure has been reported by Sakurai {\it et al}.
We are interested in the electronic state in the non-SC BLH sample, especially how the electronic state is different from that in the SC BLH sample.
This study might give a clue to know what is important for the superconductivity.   
From Co-Nuclear Quadrupole Resonance (NQR) measurements, we found a magnetic order in the non-SC BLH NCO sample. 
The transition temperature $T_M$ is 5.5 K, which is very close to the highest-$T_c$ ($T_c \sim$ 4.7 K ) in the BLH NCO system.
Our finding shows that the superconductivity is realized near the magnetic phase, suggesting strongly that the electron correlations, especially magnetic ones are essential for the superconductivity.            

Three samples we report here are listed in Table 1, together with the samples used in the previous studies.\cite{Ishida,Ihara} In Table 1, values of $T_c$, Na content $x$ and $c$-axis lattice parameter are shown.  
Details of preparation of these samples were reported elsewhere.\cite{Sakurai-Sample}
These samples are carefully characterized using an X-ray diffraction and the Na content was determined by inductively coupled plasma atomic-emission spectrometry (ICP-AES).   

\begin{table}[b]
\begin{center}
\begin{tabular}{l|rrlrr}
 No.	&\multicolumn{1}{c}{$T_{c}$}	&\multicolumn{1}{c}{$c$} 	&\multicolumn{1}{c}{$x$}	&\multicolumn{1}{c}{NQR frequency} &\multicolumn{1}{c}{Ref.}	\\ [2pt] \hline \hline
No.1   &4.7 K &19.68 {\AA} &0.348  &12.30 MHz & [7]	\\
No.2   &4.6 K &19.72 {\AA} &0.339  &12.30 MHz &	[8] \\
No.3   &2.8 K &19.57 {\AA} &0.348  &12.08 MHz &	                \\
No.4   &4.6 K &19.60 {\AA} &0.35  &12.32 MHz &	                \\
No.5   &0 K	  &19.75 {\AA} &0.331  &12.54 MHz &	              
\end{tabular}
\end{center}
\caption{List of samples used for our NMR measurements. $T_c$, $c$ and $x$ are the SC transition,  c-axis lattice constant and Na content, respectively. NQR frequency is the peak frequency of the NQR spectrum arising from the $\pm5/2\leftrightarrow \pm7/2$ transitions (see in Fig.1).}
\label{table}
\end{table}%

Table 1 signals that there is some relation between $T_{c}$ and $c$-axis constant in SC samples, as suggested previously.\cite{Milne,Ihara,Sakurai-c-axis}
It is to be noted that $x$ value is hardly correlated to $T_c$.\cite{Milne,Ihara}
We consider that the superconducting properties are determined mainly by hydrate content, which is related with the $c$-axis lattice parameter.

\begin{figure}[t]
\begin{center}
\includegraphics[width=7cm,clip]{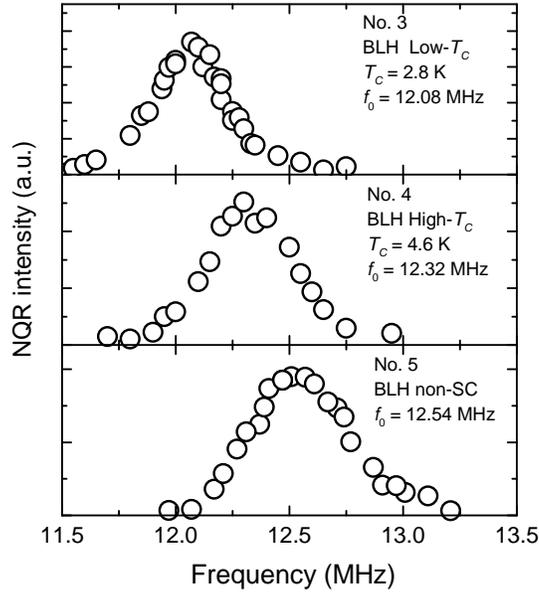}
\caption{NQR spectra arising from the $\pm 5/2 \leftrightarrow \pm 7/2$ transition in several samples with different $T_c$.}
\label{FIG1}
\end{center}
\end{figure}
Figure \ref{FIG1} shows Co-NQR spectra arising from the $\pm 5/2 \leftrightarrow \pm7/2$ transition.
We observed a single peak arising from the transition although Na ions partially occupy the Na layer. 
The double-hydrate layers screen the randomness of the Na layer, which is considered to be important for the superconductivity.   
The resonance frequency of this transition $\nu_Q$ strongly depends on samples, especially the $c$-axis lattice parameter. 
The $\nu_Q$ of the sample with $T_c \sim 4.6 $ K is nearly the same as that of previous samples with the same $T_c$, indicative of $\nu_Q$ being reproducible.
This suggests that $\nu_Q$ can be one of the good parameters to show physical properties of each sample with different $T_{c}$.

In general, $\nu_Q$ is determined by the electric field gradient (EFG) along the principal axis $V_{zz}$, (in this system, the $c$ axis is the principal axis) and the asymmetric parameter $\eta$ at the Co site.
From measurements of the resonance frequencies at other transitions, it was revealed that $\eta$ was nearly unchanged ($\eta = 0.208 \pm 0.007$) and $\nu_Q$ was changed by $V_{zz}$. 
Taking into account that the longer $c$-axis sample has the higher $\nu_Q$, we suggest that the change of $V_{zz}$ is related with the lattice distortion. 
It is noteworthy that the non-SC BLH sample has a higher $\nu_Q$ than that in other SC samples.

\begin{figure}[t]
\begin{center}
\includegraphics[width=8.5cm,clip]{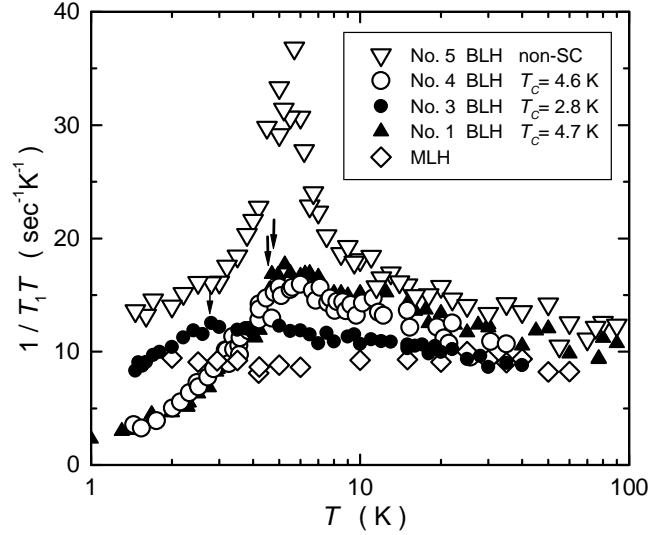}
\caption{Temperature dependence of $1/T_1T$ in various BLH NCO samples (No.3, 4,and 5) together with that in the non-SC MLH NCO sample and the SC BLH NCO sample with $T_c \sim 4.7$ K (No.1).\cite{Ishida} The arrows show $T_c$ in samples. }
\label{FIG2}
\end{center}
\end{figure}
Next we measured nuclear spin-lattice relaxation rate $1/T_{1}$ to investigate spin dynamics in these samples.
$T_1$ was measured at the NQR peaks shown in Fig.1, and can be determined by a single component in the whole temperature range.  
Figure \ref{FIG2} shows the temperature dependence of $1/T_{1}T$ of these samples, together with that of the mono-layered hydrate (MLH) NCO sample showing non-SC behavior and that of the BLH-NCO with $T_c \sim 4.7$ K (No.1) for comparison.\cite{Ishida}
The Korringa ($1/T_{1}T =$ const.) behavior was observed in the MLH sample, indicative of absence of the temperature dependent magnetic fluctuations.
In SC samples, the value of $1/T_{1}T$ increases with decreasing temperature from 50 K to $T_c$, and the value of $1/T_1T$ at $T_c$ is larger in the higher-$T_c$ sample.
These behaviors are in good agreement with the previous results.\cite{Ishida,Ihara}
The remarkable finding is that $1/T_1T$ of the non-SC BLH sample shows a prominent peak at $T_M \sim 5.5$ K, suggestive of occurrence of a magnetic ordering.
$T_M \sim$ 5.5 K is close to the SC transition temperature $T_c$. 
$1/T_1T$ above $T_M$ is larger than that in other SC samples, indicating that magnetic fluctuations in the non-SC BLH sample are stronger than those in the high-$T_c$ sample.
It should be noted that the temperature dependence of $1/T_1T$ is the same for the non-SC sample and the high-$T_c$ sample in the temperature range of 10 - 100 K.
These experimental results suggest strongly that the unconventional superconductivity in the BLH NCO occurs in the vicinity of the magnetic ordering, and that the magnetic fluctuations in the high-$T_c$ sample are close to the magnetic instability.

\begin{figure}[t]
\begin{center}
\includegraphics[width=8cm,clip]{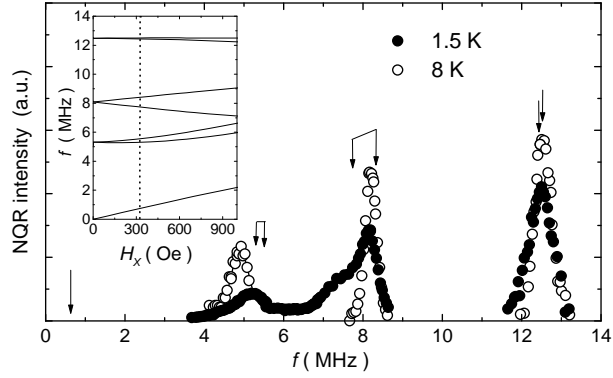}
\caption{Co NQR spectra of 1.5 K (black) and 8 K (white). 
The transition of 8.2 MHz is largely affected by the internal field. 
The inset shows the dependence of transition frequencies arising from all transitions with respect to the change of the internal field along the CoO$_2$ plane $H_x$. 
The peaks at 1.5 K are interpreted by $H_x \sim$ 300 Oe. The calculated peaks are shown by arrows in the main figure.}
\label{FIG3}
\end{center}
\end{figure}

In order to know the magnetically-ordered state, NQR spectra arising from all the transitions are obtained.
Figure 3 shows the NQR spectra above (8 K) and below (1.5 K) $T_M$. 
Three single peaks corresponding to the $\pm 1/2 \leftrightarrow \pm 3/2$ (4.90 MHz), $\pm 3/2 \leftrightarrow \pm 5/2$ (8.20 MHz), and $\pm 5/2 \leftrightarrow \pm 7/2$ (12.54 MHz) transitions are observed at 8 K. 
At 1.5 K below $T_M$, the peaks at 4.90 MHz and 12.54 MHz becomes slightly broader, whereas the peak at 8.20 MHz splits into two peaks asymmetrically.
It seems that the effect of the internal field appears mainly in the $\pm 3/2 \leftrightarrow \pm 5/2$ transition.

In general, the NQR Hamiltonian with an internal field $\overrightarrow{H} = (H_x, 0, H_z)$ is expressed as,
\[\mathcal{H} = \frac{\hbar\nu_z}{6} \{3I_{z}^{2}+I^{2}+\eta (I_{+}^{2}+I_{-}^{2})\} +\gamma \hbar (I_{x}H_{x}+I_{z}H_{z}) \]
Here $\nu_z$ is defined as $\nu_z =3 eQV_{zz}/3I(2I-1)$ with $Q$ being the nuclear quadrupolar moment, and the asymmetric parameter $\eta$ is $|\nu_x-\nu_y|/\nu_z$.
If $H_z$ were dominant, all transitions would be affected by the operator $I_z$. 
Obviously this is not the case.
The inset shows calculated curves of three-transition frequencies with respect to the change of an internal field along the plane $H_x$, in which the experimental data of $\nu_{z} \sim 4.2$ MHz and $\eta \sim 0.203$ evaluated from the NQR spectra above $T_{M}$ are used.
As seen in the inset, the transition at 8.20 MHz is largely affected by $H_{x}$.
The NQR spectrum peaks at 1.5 K are consistently fitted by $H_x \sim$ 300 Oe as shown in the inset.
The asymmetric spectra at $\sim$ 8.2 MHz are interpreted by the difference of the transition probability of two split peaks.
From the internal field at the Co site, the ordered moment is evaluated as small as 0.015 $\mu_B$ by using the hyperfine coupling constant being $\sim 20$ kOe/$\mu_{B}$ derived from the $K$ - $\chi$ plot in the normal state.\cite{Michioka}

\begin{figure}[t]
\begin{center}
\includegraphics[width=7cm,clip]{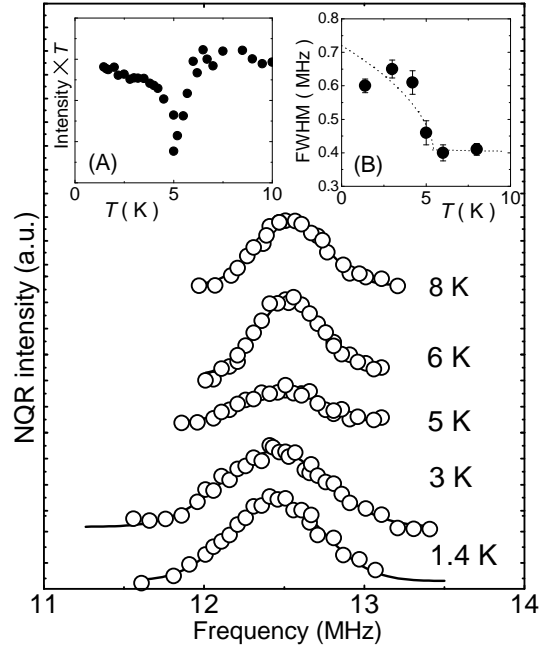}
\caption{Co-NQR spectra from the $\pm 7/2 \leftrightarrow \pm 5/2$ transition at various temperatures in the non-SC BLH sample. The inset (a) and (b) show the temperature dependence of the integrated intensity and the full width at half maximum (FWHM) of the spectrum, respectively. The dotted curves in the inset shows the mean-field-behavior, $\sim (1-(T/T_M))^{1/2}$.  }
\label{FIG4}
\end{center}
\end{figure}

To know the volume fraction of the ordered state and the temperature dependence of the ordered moments, the NQR spectra from the $\pm5/2 \leftrightarrow \pm7/2$ transition were obtained at various temperatures.
Figure 4 shows the NQR spectra below 8 K.
The insets (a) and (b) are the temperature dependence of the integrated intensity and the full width at half maximum (FWHM) of the NQR spectra, respectively.
Due to the divergence of $1/T_{1}T$, the integrated intensity shows a minimum at 5.5 K, below which FWHM becomes broader. 
The temperature dependence of FWHM shows the mean-field-type behavior.
It should be noted that the integrated intensity of the spectrum recovers totally at 1.5 K. 
This experimental fact excludes the possibility that some magnetic fraction is missing due to the occurrence of the large internal field, 
rather suggests that the weak magnetism in which the ordered moments are approximately 10$^{-2} \mu_B$ occurs in the entire region of the non-SC BLH sample.

\begin{figure}[t]
\begin{center}
\includegraphics[width=8cm,clip]{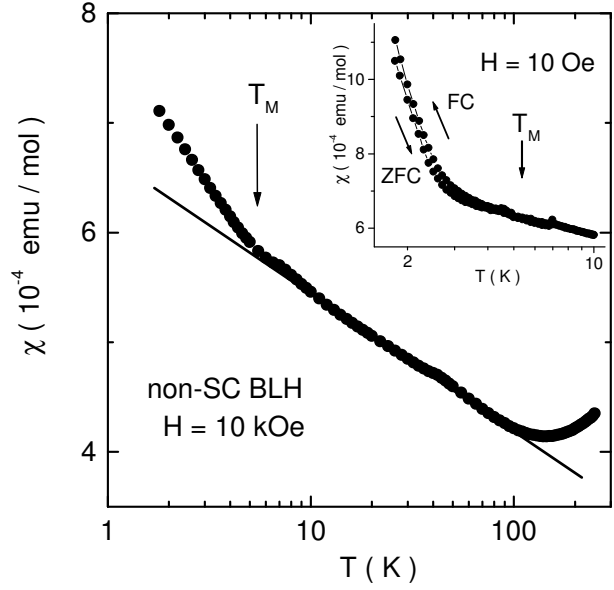}
\caption{Temperature dependence of magnetic susceptibility $\chi_{\rm bulk}$ in the non-SC BLH samples measured in $\sim 20$ kOe. Horizontal axis is a logarithmic scale. Anomaly is observed at 5.5 K, below which $\chi_{\rm bulk}$ shows an upward increase. The inset shows $\chi_{\rm bulk}$ measured in a small field of 10 Oe.}
\label{FIG5}
\end{center}
\end{figure}

In order to consider what kind of a magnetic order occurs, we discuss the temperature dependence of the bulk susceptibility $\chi_{\rm bulk}$ of the non-SC BLH sample.
Figure 5 shows the temperature dependence of $\chi_{\rm bulk}$ of the powder sample measured at 10 kOe.   
A clear anomaly is observed at $T_M \sim$ 5.5 K, below which $\chi_{\rm bulk}$ shows an upward increase below $T_M$.
In typical antiferromagnetic (AFM) compounds $\chi_{\rm bulk}$ shows a peak at $T_M$, while, $\chi_{\rm bulk}$ diverges in  ferromagnetic (FM) compounds. 
The inset of Fig. 5 shows the temperature dependence of $\chi_{\rm bulk}$ under a small field of 10 Oe. 
Tiny hysteresis was observed between zero-field-cooled (ZFC) and field-cooled (FC) measurements. 
This behavior might exclude the occurrence of the  spin-glass and/or cluster-glass transition since a large hysteresis between FC and ZFC would be observed due to a spin freezing in these cases. 
The upward increase observed in the non-SC BLH sample can not be interpreted by ordinary magnetic anomaly, and therefore deserves to be clarified by further experiments.

\begin{figure}[t]
\begin{center}
\includegraphics[width=8cm,clip]{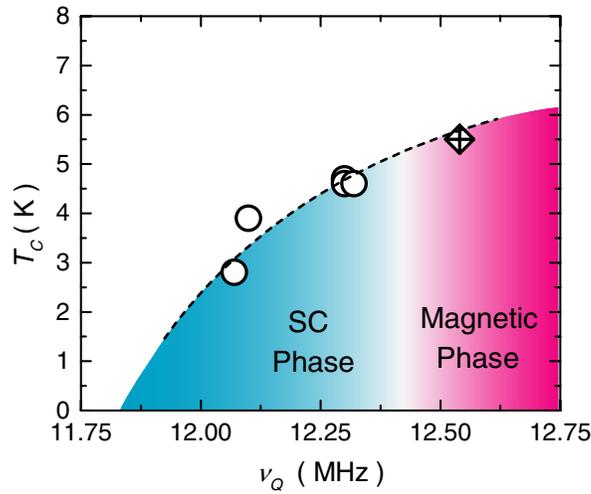}
\caption{Proposed phase diagram of Na$_{x}$CoO$_{2}\cdot y$H$_{2}$O.
The vertical and horizontal axes are transition temperature and resonance frequency $\nu_Q$, respectively.
The boundary between SC and magnetic phases seems to be present around 12.4 MHz. The data in Ref. [8] are also plotted.}
\label{FIG6}
\end{center}
\end{figure}

Finally, we propose a possible phase diagram of Na$_{x}$CoO$_{2}\cdot y$H$_{2}$O using the experimental value of $\nu_Q$ as a tuning physical parameter in Fig.6.
It is obvious that the magnetic phase is adjacent to the superconducting phase, although the relation and boundary between the two phases are not clear yet.
If the superconductivity were induced by the electron-phonon mechanism, the superconductivity would be suppressed by strong magnetic fluctuations close to the magnetic instability.
Instead, taking into account that the SC transition temperature smoothly changes into the magnetic-ordering temperature, we suggest that the magnetic fluctuations which induce the magnetic ordering might be related with the superconductivity. 

Now, we consider physical meaning of the tuning parameter $\nu_Q$. 
As discussed above, $\nu_Q$ is determined by the EFG along the $c$ axis, $V_{zz}$ which is related with the crystal distortion of the CoO$_{2}$ block layers.
In addition, it was reported from the neutron-scattering experiment that the strong inverse correlation between the CoO$_2$-layer thickness and $T_c$ ($T_c$ increases with decreasing the thickness).\cite{Lynn}   
Taking into account these experimental results, we consider that a decrease of the layer thickness, which makes $\nu_Q$ larger, corresponds to a compression of the tilted octahedron along the $c$ axis.
Such a distortion makes the three degenerated $t_{2g}$ orbitals futher split into one $a_{1g}$ and two $e'_{g}$ orbitals.
According to theoretical calculations, hole-character Fermi surface (FS) of the six hole pockets, which are formed by the $e'_{g}$ orbitals ($e'_g$-FS), becomes larger when the compression along the $c$ axis becomes larger.\cite{Ikeda-private} 
Therefore, we consider that the change of the parameter $\nu_Q$ corresponds to change of the $e'_g$-FS volume.
It was reported from the theoretical calculation that the static spin susceptibility around the $\Gamma$ point is enhanced and the spin-triplet superconductivity with $p$- or $f$-wave character is stabilized when the $e'_g$-FS volume becomes larger.\cite{Yanase}
In this meaning, the spin fluctuations, which are enhanced when $\nu_Q$ becomes larger, might have a ferromagnetic character.
However, a crucial discrepancy was reported: spin-singlet superconductivity was suggested by Co-Knight shift measurements.\cite{Kobayashi}
One of the most important questions to be answered is how the spin fluctuations around the $\Gamma$ point, which is close to the magnetic instability, can coexist with spin-singlet superconductivity.
Further experiments are still needed to clarify the SC and spin-fluctuation nature in the BLH NCO system.\cite{Comment} 

In conclusion, we found the weak magnetic order below $T_M \sim$ 5.5 K in a non-SC Na$_{x}$CoO$_{2}\cdot y$H$_{2}$O sample with the same structure as SC samples.
The ordered moment is approximately 0.015 $\mu_B$, which resides in the CoO$_2$ plane.
The unusual upward increase was observed in $\chi_{\rm bulk}$ below $T_M$, which are different from behaviors observed below an ordinary magnetic transition.    
From NQR measurements on various BLH NCO samples with different $T_c$, we suggest that $\nu_Q$ could be a tuning parameter of the ground state in this system, and that the occurrence of the superconductivity might be related to the $c$-axis distortion of the tilted CoO$_6$ octahedron. 

We thank H.~Yaguchi, Y.~Maeno, and S.~Nakatsuji for experimental support and valuable discussions. We also thank H.~Ikeda, S.~Fujimoto, K.~Yamada, Y.~Yanase, M. Mochizuki, M.~Ogata and W.~Higemoto for valuable discussions.
This work was supported by CREST of JST, by the 21 COE program on ``Center for Diversity and Universality in Physics'' from MEXT of Japan, and by Grants-in-Aid for Scientific Research from JSPS and MEXT.

\end{document}